\documentclass[prl,preprint,preprintnumbers,amsmath,amssymb,superscriptaddress]{revtex4}
\usepackage{graphicx}% Include figure files
\usepackage{epsfig}
\usepackage{dcolumn} %decimal point spacing table
\usepackage{bm} %bold math mode
\begin{document}

\title{Potential barrier lowering and electrical transport at the LaAlO$_{3}$/SrTiO$_{3}$ heterointerface}

\author{Franklin J. Wong}
\affiliation{Department of Materials Science and Engineering, University of California, Berkeley, CA 94720}
\author{Miaofang Chi}
\affiliation{Department of Chemical Engineering and Materials Science, University of California, Davis, CA, 95616}
\affiliation{Lawrence Livermore National Laboratory, Livermore, CA, 94550}
\author{Rajesh V. Chopdekar}
\affiliation{Department of Materials Science and Engineering, University of California, Berkeley, CA 94720}
\affiliation{School of Applied and Engineering Physics, Cornell University, Ithaca, NY 14853}
\author{Brittany B. Nelson-Cheeseman}
\affiliation{Department of Materials Science and Engineering, University of California, Berkeley, CA 94720}
\author{Nigel D. Browning}
\affiliation{Department of Chemical Engineering and Materials Science, University of California, Davis, CA, 95616}
\affiliation{Lawrence Livermore National Laboratory, Livermore, CA, 94550}
\author{Yuri Suzuki}
\affiliation{Department of Materials Science and Engineering, University of California, Berkeley, CA 94720}

\date{\today}

\begin{abstract}
Using a combination of vertical transport measurements across and lateral transport measurements along the LaAlO$_{3}$/SrTiO$_{3}$ heterointerface, we demonstrate that significant potential barrier lowering and band bending are the cause of interfacial metallicity.  Barrier lowering and enhanced band bending extends over 2.5 nm into LaAlO$_{3}$ as well as SrTiO$_{3}$.  We explain origins of high-temperature carrier saturation, lower carrier concentration, and higher mobility in the sample with the thinnest LaAlO$_{3}$ film on a SrTiO$_{3}$ substrate.  Lateral transport results suggest that parasitic interface scattering centers limit the low-temperature lateral electron mobility of the metallic channel.   
 
\end{abstract}

\maketitle

Interfacial phenomena form the basis for modern-day devices and continue to be an exciting area in condensed matter research.  The engineering of two-dimensional electron gases and the discovery of new physical phenomena such as the quantum Hall effect \cite{klitzing} have been realized at conventional semiconductor interfaces.  Advances in oxide thin-film fabrication have enabled the synthesis of atomically precise oxide interfaces and hence allowed for controlled investigation of interfacial phenomena in these materials.  With the rich variety of functionalities exhibited by transition-metal oxides, a wide array of novel properties may be achieved at oxide heterointerfaces.  An exemplary study is the discovery of metallicity at the interface of two band insulators, LaAlO$_{3}$ (LAO) and SrTiO$_{3}$ (STO) \cite{hwang}, which has stimulated many subsequent experimental \cite{muller,thiel,schneider,huijben,siemons,koster,fert,brinkman,kalabukhov,reyren,levy} as well as theoretical studies \cite{levy,pentcheva,gemming,park}.  However, there is still intense debate on the origin of metallicity, specifically whether it arises from electronic reconstruction \cite{hwang,muller} or oxygen vacancies \cite{koster,fert,kalabukhov}.   

In this Letter, we demonstrate that metallicity observed in our LAO/STO heterostructures can be attributed to potential barrier lowering and band bending at the LAO/STO interface.  With vertical transport measurements, we show that the thickness of the metallic region extends to at least several nanometers and is not confined to the order of a unit cell as has been theoretically predicted \cite{pentcheva,gemming,park}.  We will argue that oxygen vacancies cannot be the sole source of metallicity.  Lateral transport measurements of LAO films on STO substrates indicate carrier saturation at high temperatures and higher low-temperature mobility values in the thinnest LAO film, features that we show to be consistent with charge transfer-induced metallicity.   

We used pulsed laser deposition to deposit two types of samples: (1) vertical stacks composed of SrRuO$_{3}$ metal electrodes sandwiching thin LAO and/or STO layers and (2) LAO films of varying thickness on TiO$_{2}$-terminated (100) STO substrates.  The vertical stacks form a tunnel junction geometry: TiO$_{2}$-terminated(100) STO substrate // SrRuO$_{3}$(60 nm) / LAO (2.5 nm) / STO (2.5 nm) / SrRuO$_{3}$ (40 nm).  This stack will be referred to as VS1.  Since SrRuO$_{3}$ films strongly prefer SrO surface termination \cite{eom} on TiO$_{2}$-terminated STO substrates, charge neutrality considerations dictate that the LAO/STO interface is a (LaO)$^{+}$/TiO$_{2}$ \textit{n}-type interface.  As reference structures, tunnel junctions with SrRuO$_{3}$ electrodes using only either a 5 nm LAO or 5 nm STO film as the barrier layer were fabricated and will be referred to as VS2 and VS3 respectively. The bottom SrRuO$_{3}$ electrode was deposited at 700 $^{\circ}$C, 1.4 J/cm$^{2}$, 2 Hz pulse rate, and 60 mTorr O$_{2}$.  The top electrode was deposited at 635 $^{\circ}$C, 1.4 J/cm$^{2}$, 4 Hz, and 60 mTorr O$_{2}$.  The LAO and STO layers were deposited at 700 $^{\circ}$C, 1.4 J/cm$^{2}$, 2 Hz, and 2$\times$10$^{-5} $Torr O$_{2}$. The entire stack was annealed at 600 $^{\circ}$C for seven minutes and then at 400 $^{\circ}$C for one hour in $\approx$300 Torr O$_{2}$.  Single LAO films were grown on TiO$_{2}$-terminated (100) STO substrates at 700 $^{\circ}$C, 1.4 J/cm$^{2}$, 2 Hz, and 2$\times$10$^{-5}$ Torr O$_{2}$. We will refer to the single-layer samples of 2.5, 6.5, and 14 nm-thick LAO films as LAO1, LAO2, and LAO3 respectively.   

\begin{figure}
\includegraphics[width = 3.37 in]{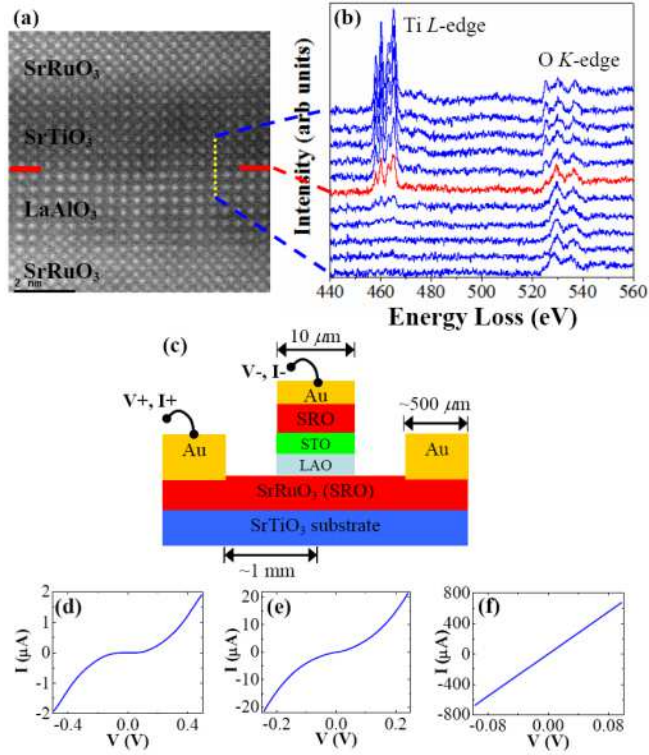}% Here is how to import EPS art
\caption{\label{fig_STEM}(Color online) (a) STEM image of the SrRuO$_{3}$-based vertical stack VS1 along the [001] zone axis.  This is a representative image of all regions in the sample. (b) A corresponding EELS linescan revealing that interdiffusion at the LAO/STO interface is limited. (c) Schematic of the vertical stack structure.  Figure not drawn to scale.  IV curves of vertical transport measurements performed on stacks with (d) only LAO (VS2), (e) only STO (VS3), and (f) a composite double layer of LAO and STO in between two SrRuO$_{3}$ electrodes (VS1).}
\end{figure}

A combination of electron microscopy and X-ray diffraction confirmed the excellent crystallinity of all the samples.  Scanning Transmission Electron Microscopy (STEM) reveals sharp interfaces in the multilayered heterostructure VS1 throughout the entire sample.  Figure~\ref{fig_STEM}a is a representative STEM image of VS1.  Electron energy loss spectroscopy (EELS) analysis, using a 2 $\AA$ probe, showed that cation interdiffusion was limited to one unit cell at the LAO/STO heterointerface.  Figure~\ref{fig_STEM}b is a representative atomic-resolution EELS linescan of Ti \textit{L}- and O \textit{K}-edges in half unit cell steps across the LAO/STO interface, confirming atomic sharpness.  

Two-point vertical transport measurements were performed at 5 K on the vertical stack junctions, with the current flowing from one SrRuO$_{3}$ electrode across to the other SrRuO$_{3}$ electrode.  Figure~\ref{fig_STEM}c is a schematic of the vertical structure.  Figures~\ref{fig_STEM}d-f show that VS2 and VS3 (4 $\mu$m $\times$ 4 $\mu$m areal size), containing either only a LAO or only a STO barrier layer, exhibit non-linear current-voltage (IV) curves, as expected for tunneling conductance across metal-insulator-metal junctions.  The stack with composite inter-layers, VS1 (10 $\mu$m $\times$ 10 $\mu$m areal size), exhibits a linear IV curve, indicative of ohmic conduction.  Ohmic conduction in VS1 but not in VS2 or VS3 suggests that enhanced band bending induced by interface states effectively thins and lowers the potential barrier of the LAO/STO interface.  We deduce that band bending extends to at least six unit cells on both sides of the LAO/STO interface.  Therefore, although the heterointerface is atomically sharp, as seen in Figure~\ref{fig_STEM}a, it is not electronically sharp.  

\begin{figure}
\includegraphics[width = 3.35 in]{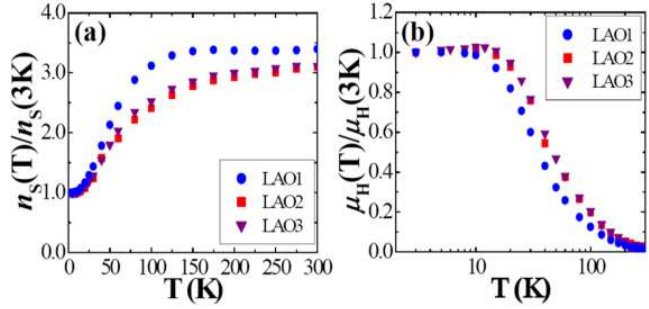}% Here is how to import EPS art
\caption{\label{fig_concentrationandmobility}(Color online) (a) Normalized sheet carrier concentration $\textit{n}_{S}$(T)/$\textit{n}_{S}$(3K) and (b) normalized Hall mobility $\textit{$\mu$}_{H}$(T)/$\textit{$\mu$}_{H}$(3K) curves of LAO1 to LAO3.}  
\end{figure}

In order to probe lateral transport along the interface, four-point van der Pauw sheet resistance and Hall effect measurements were performed on LAO1 to LAO3.  Table~\ref{table_ParametersTable} summarizes some of their lateral transport properties.  The sheet carrier concentration ($\textit{n}_{S}$) shows no scaling with film thickness, suggesting the measured conductivity is likely confined to the interface.  Plotted in Figures~\ref{fig_concentrationandmobility}a and b are the $\textit{n}_{S}$ and Hall mobility ($\textit{$\mu$}_{H}$) values normalized by their values measured at 3K - i.e. $\textit{n}_{S}$(T)/$\textit{n}_{S}$(3K) and  $\textit{$\mu$}_{H}$(T)/$\textit{$\mu$}_{H}$(3K).  The normalized curves emphasize the otherwise subtle differences as a function of LAO film thickness.  The $\textit{$\mu$}_{H}$ values in all of the samples are similar in magnitude at room temperature, suggesting a common scattering mechanism such as carrier interaction with optical phonons.  Low-temperature mobility is likely to be limited by carrier scattering at the heterointerface.  The width of the temperature range in which the mobility saturates is an indirect measure of the strength of electron coupling to the interface.  LAO1 has the narrowest saturated mobility region (Figure~\ref{fig_concentrationandmobility}b) as well as the largest low-temperature mobility values (Table~\ref{table_ParametersTable}), telling us that compared to LAO2 and LAO3, there is relatively weaker scattering of carriers by the interface.  The similarities in both $\textit{n}_{S}$(T)/$\textit{n}_{S}$(3K) and  $\textit{$\mu$}_{H}$(T)/$\textit{$\mu$}_{H}$(3K) of LAO2 and LAO3 imply that beyond a certain film thickness, band bending at the interface equilibrates, and electrical transport behaviors remain approximately constant.  

\begin{table}
\includegraphics[width = 3.3 in]{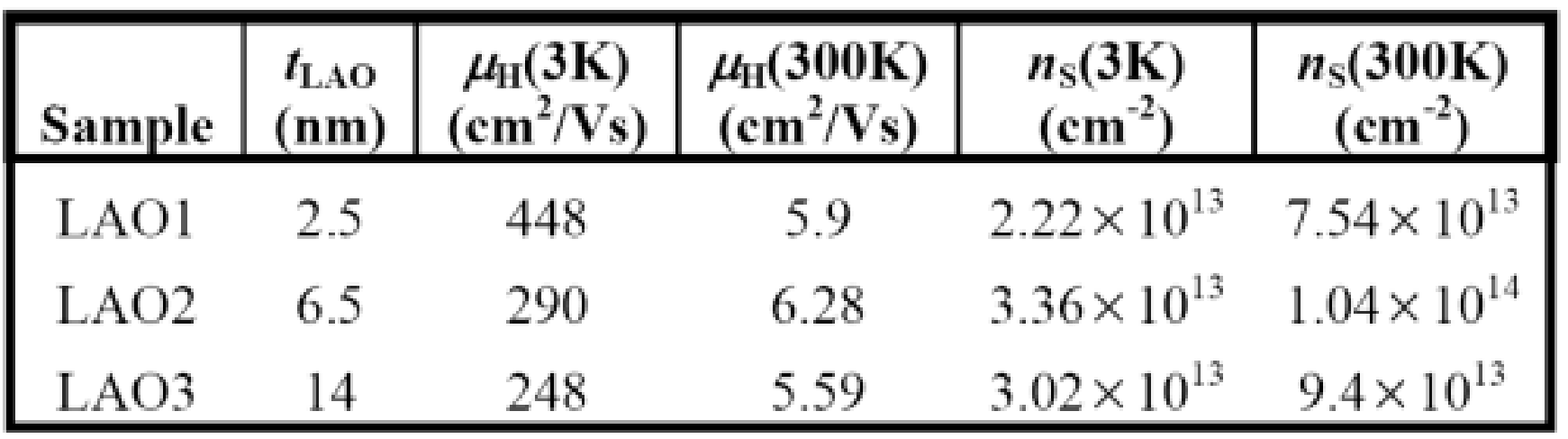}% Here is how to import EPS art
\caption{\label{table_ParametersTable}(Color online) Selected parameters of LAO1 to LAO3, where \textit{t}$_{LAO}$ is the thickness of the LAO film.}
\end{table}

Oxygen vacancies in STO have been argued to be the source of metallicity in LAO/STO heterostructures \cite{koster,fert,kalabukhov}, but we believe that this is not the case in our samples.  The tunneling behavior observed in VS2 and VS3 supports our claim that our LAO and STO layers are well oxygenated and that interface effects may be the origin of barrier lowering and interfacial metallicity.  The magnitudes of the low-temperature $\textit{$\mu$}_{H}$ values of LAO1 to LAO3 (Table~\ref{table_ParametersTable}) match those of bulk STO single crystals lightly doped with oxygen vacancies \cite{lee,brehner} and are markedly lower than those of  heavily doped STO samples with typical low-temperature $\textit{$\mu$}_{H}$ values of up to 5000 cm$^{2}$/Vs \cite{tufte}.  However, lightly doped STO crystals show clear carrier freeze-out \cite{lee,brehner}, which is in stark contrast with the $\textit{n}_{S}$(T)/$\textit{n}_{S}$(3K) behavior of our samples.  Furthermore, conduction through the substrate is at odds with the observation of an insulating \textit{p}-type interface formed by (AlO$_{2}$)$^{-}$ and (SrO)$^{0}$ planes, reported by Ohtomo and Hwang \cite{hwang}.  We have also deposited homoepitaxial STO films using the same conditions as in LAO1 to LAO3.  These samples are too insulating to be measured electrically, thus indicating that the growth conditions in themselves do not cause metallicity in STO substrates.   
		
Together the vertical and lateral transport measurements show that the LAO/STO interface is characterized by the following features: (a) The formation of the heterointerface lowers the potential barrier of the LAO/STO interface for electron tunneling.
(b) The largest carrier concentration changes in LAO1 to LAO3 occurs between 20 and 100 K, correlating with the strong temperature dependence of the dielectric constant of STO \cite{neville}. (c) Carrier concentration is lowest in the thinnest LAO film on STO (LAO1), which also shows carrier saturation at high temperatures. (d) Electron mobility values for all samples are low compared to those of degenerately doped STO single crystals at low temperatures. 
     
There are a number of mechanisms that may explain our experimental observations.  They include carrier introduction via charge transfer at the polar LAO/STO interface, lattice deformation, and interface chemical bonding effects.  Now, we will focus one of the possible mechanisms in greater detail - carrier introduction via charge transfer at the polar interface of two nominally undoped insulators.  We will assume a conduction band offset  (E$_{C}$) of about 2.3 eV and valence band offset (E$_{V}$) of about 0.1 eV, as given by recent band offset calculations \cite{demkov,albina}.  Additionally, since both LAO and STO are undoped, their Fermi levels are assigned near mid-gap. 

\begin{figure}
\includegraphics[width = 3.375 in]{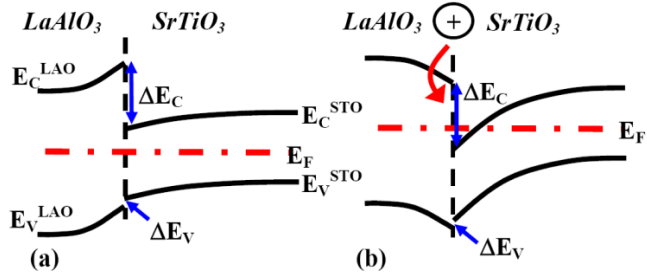}% Here is how to import EPS art
\caption{\label{fig_schematicalignment}(Color online) (a) Schematic band alignment of LAO and STO without charged interface states.  Band bending occurs in order to equilibrate the Fermi energy level. (b) Schematic band alignment of LAO and STO with the inclusion of positive interface charges, resulting in the increased downward bending of all band edges.}
\end{figure}

In the case where there are no charged interface states, the band alignment would be as shown in Figure~\ref{fig_schematicalignment}a, and there would not be a metallic channel.  Metallic conduction can be achieved if there are positively charged interface states that act to pull the electronic bands downwards in energy, leading to the STO CB crossing the Fermi level, as shown in Figure~\ref{fig_schematicalignment}b.  Despite its simplicity, this band alignment description can account for the interfacial metallicity observed between two undoped insulators.  We will now discuss a possible source of such positive interface states.   

After the creation of the polar LAO-STO heterointerface, there is a divergence of potential energy, and the electronic bands of LAO continue to gain energy in layers farther from the interface \cite{muller}.  Beyond a critical thickness, the LAO valence band (VB) becomes higher in energy than the STO CB at the interface, and electrons can then tunnel from the LAO VB to the STO CB.  We hypothesize that LAO sources interfacial electrons to the STO side, and the electrons form a metallic electron channel.  With the approximation of a charge density of +/- one unit charge per half unit cell, i.e. (LaO)$^{+}$ or (AlO$_{2}$)$^{-}$, the dielectric constant of LAO to be 25, and the bandgap of STO to be 3.2 eV, the reported insulator-metal transition critical thickness of four LAO unit cells \cite{thiel} can be reproduced.  

In this description, there are holes in LAO and electrons in STO.  In all of our samples, the effective charge carriers are electron-like.  While Hall effect measurements cannot rule out the possibility of a hole current contributing to the conduction, there are several mechanisms that can trap holes in LAO: (1) an on-site repulsion energy in the valence O-2\textit{p} bands of LAO \cite{pentcheva}, (2) strong electron-lattice interactions, or (3) negative \textit{U} pairing centers, as in many conventional semiconductors \cite{mooney}.  It has been predicted that polar LAO surfaces are susceptible to the accumulation of carriers \cite{demkov,lanier,tang}, but to the authors' knowledge, there have not been any reports of \textit{n}- or \textit{p}-type metallic conductivity in LAO.  Therefore, the holes in LAO are likely to be immobile and act as localized positively charged interface states, i.e. states above the Fermi level necessary to change the band alignment from Figure~\ref{fig_schematicalignment}a to b.  The interface potential barrier is therefore lowered.   

The conincident temperature dependence of $\textit{n}_{S}$(T)/$\textit{n}_{S}$(3K) in LAO1 to LAO3 and the dielectric constant of STO now appears to be linked though interface band bending.  However, the dielectric function near the interface is complicated by the strong electric fields, mobile carriers, lattice distortions, and other interfacial effects.  Quantitative effects on the dielectric response of the heterointerface are not the focus of our qualitative description. 

The apparent carrier saturation in LAO1 supports our claim that the conduction electrons originate from the VB of LAO.  In the sample with the thinnest LAO film, the supply of electrons is depleted at high temperatures.  Therefore, together with the vertical transport results, we estimate the physical length of band bending in LAO to be between 2.5 and 6.5 nm.  We infer that the LAO films in LAO2 and LAO3 are thicker than the equilibrium width of band bending on the LAO side.  

In LAO/STO heterostructures, the large concentration of positively charged holes in the LAO side of the interface can strongly scatter electrons in the interfacial channel through Coulomb attraction, thus explaining the comparatively low mobility values of LAO1 to LAO3 (Table~\ref{table_ParametersTable}).  In addition, the strain caused by the 3\% lattice mismatch between LAO and STO is likely to induce lattice distortions near the interface.  Indeed, theoretical calculations have predicted distortions, possibly ferroelectric-like, in STO near the LAO/STO interface \cite{gemming,park}.  While screening by mobile electrons precludes the prospect of long-range ferroelectric ordering, electric dipoles formed by lattice deformation would dramatically degrade electron mobility.  Though intrinsic charge transfer offers a crude form of modulation doping, parasitic interface effects ultimately undermine any potential of mobility enhancement.

We would like to reiterate that other effects relating to interface bonding and/or lattice deformation can induce similar effects in this and related heterostructures.  Band offset and bending of the LAO/STO interface warrant further attention before engineering control of this and similar heterostructures can be attained.   

In summary, we have shown that there is significant band bending on both sides of the LAO/STO heterointerface.  We have concluded that deposition conditions alone cannot cause metallic conductivity in our samples.  Our experiments provide strong evidence for band bending, potential barrier lowering and thinning, as well as interfacial metallicity induced by charge transfer.  Although intrinsic charge transfer brings about interfacial metallicity, the lack of extrinsic control limits the electron mobility.   
 
This work was supported by the U.S. Department of Energy under Grant Nos. DE-AC02-05CH11231 and DE-FG02-03ER46057.  We would like to thank Dr. Kin Man Yu for Rutherford Backscattering measurements.  Electron microscopy and sample preparation were performed at National Center for Electron Microscopy, Lawrenece Berkeley National Laboratory.  A 200kV FEI F20 UT Tecnai was used to take STEM images and EELS spectra.  Patterning of vertical stack heterostructures were performed in the Microlab at University of California, Berkeley.  We would like to acknowledge and thank the staff of both facilities for their technical assistance.

\bibliography{SuzukiCondMattSept2008}% Produces the bibliography via BibTeX.

\end{document}